\documentclass[aps,prb,10pt,twocolumn,showpacs,floatfix,superscriptaddress,amsmath,amssymb]{revtex4-1}

\usepackage{graphicx}% Include figure files
\usepackage{dcolumn}% Align table columns on decimal point
\usepackage{bm}% bold math
\usepackage{pstricks}% bold math

\def\nn{\nonumber}
\newcommand{\ba}{\begin{eqnarray}}
\newcommand{\ea}{\end{eqnarray}}
\def\be{\begin{equation}}
\def\ee{\end{equation}}

\def\prl#1#2#3{Phys.\ Rev.\ Lett.\ {\bf #1}, #2 (#3)}
\def\prb#1#2#3{Phys.\ Rev.\ B {\bf #1}, #2 (#3)}

\def\bimno{Bi$_3$Mn$_4$O$_{12}$(NO$_3$)}

\begin{document}

\title{Quantum phases in the frustrated Heisenberg model
on the bilayer honeycomb lattice}

\author{Hao Zhang}
\email{zhanghao@issp.u-tokyo.ac.jp}
\affiliation{Institute for Solid State Physics, University of Tokyo, Kashiwa, Chiba 277-8581, Japan}

\author{ M.\ Arlego}
\affiliation{IFLP - CONICET, Departamento de F\'isica, Universidad Nacional de La Plata,
C.C.\ 67, 1900 La Plata, Argentina.}

\author{ C. A.\ Lamas}
\email{lamas@fisica.unlp.edu.ar}
\affiliation{IFLP - CONICET, Departamento de F\'isica, Universidad Nacional de La Plata,
C.C.\ 67, 1900 La Plata, Argentina.}

\begin{abstract}
We use a combination of analytical and numerical techniques to study
the phase diagram of the frustrated Heisenberg model on the bilayer
honeycomb lattice.
Using the Schwinger boson description
of the spin operators followed by a mean field decoupling, the
magnetic phase diagram is studied as a function of the frustration
coupling $J_{2}$ and the interlayer coupling $J_{\bot}$.

The presence of both magnetically ordered and disordered phases is
investigated by means of the evaluation of ground-state energy, spin
gap, local magnetization and spin-spin correlations. We observe a
phase with a spin gap and short range N\'eel correlations that
survives for non-zero next-nearest-neighbor interaction and
interlayer coupling. Furthermore, we detect signatures of a
reentrant behavior in the melting of N\'eel phase and symmetry
restoring when the system undergoes a transition from an on-layer
nematic valence bond crystal phase to an interlayer valence bond
crystal phase. We complement our work with exact diagonalization on
small clusters and dimer-series expansion calculations, together
with a linear spin wave approach to study the phase diagram as a
function of the spin $S$, the frustration and the interlayer
couplings.
\end{abstract}

\maketitle

%-------------------------------------------------------------------

\section{Introduction}

The study of the possible disordered ground states on the honeycomb lattice has received a
 great interest in the last years.
 The interest is focused mainly on the existence of quantum spin liquids in quantum antiferromagnets\cite{Anderson,Balents_Nature,ML_2005,Oshikawa_2013,LM_2011}.
 Recently, possible quantum disordered phases have been reported in the phase diagram corresponding to the single layer honeycomb Heisenberg
 model\cite{Wang,Cabra_honeycomb_prb,Cabra_honeycomb_2,Zhang_PRB_2013,Mulder,Mosadeq,Albuquerque,Reuther,Farnell,Bishop_2012,Li_2012_honeyJ1-J2-J3,Bishop_2013,Clark,Mezzacapo,Ganesh_PRL_2013,Zhu_PRL_2013,Fisher_2013,Albuquerque_capponi_2012}.
 From the theoretical point of view, it is interesting to study the influence of an interlayer coupling
 in the stabilization of these disordered phases.
 In particular in the bilayer models, the ground state corresponding to very large values of the interlayer
 couplings should be a dimer product state. For unfrustrated models a transition between
 the N\'eel phase and the dimer phase is expected to obtain as the interlayer coupling is increased.
 This ``melting'' of N\'eel order can be studied as a function of the frustration in each layer.
 By contrast in the frustrated case, the  system might go from a nonmagnetic nematic phase to a
 dimer product state as the interlayer coupling is increased.

From the experimental side, a very exciting progress on the
bismuth oxynitrate, {\bimno}, was obtained by Smirnova
\textit{et al.}\cite{BiMnO}. The magnetic susceptibility data indicate two-dimensional magnetism.
Despite the large AF Weiss constant of -257K, no long-range ordering
was observed down to 0.4K, which suggests a nonmagnetic ground
state\cite{BiMnO,ESR,expnew2,Azuma}.
In this compound the Mn$^{4+}$ ions form a honeycomb lattice without
any distortion. Two layers of such honeycomb lattices are separated
by bismuth atoms, forming a bilayer structure with these bilayers
separated by a large distance. Thus, the appropriate geometry to describe
its magnetic properties is the bilayer honeycomb lattice.
The magnetic exchange coupling constants have been calculated using
a density functional theory, which shows that the dominant
interactions are the intra-layer nearest-neighbor interaction $J_1$
and the effective interlayer interaction $J_{\bot}$\cite{Kandpal}.

\begin{center}
\begin{figure}[t!]
\includegraphics[width=0.9\columnwidth]{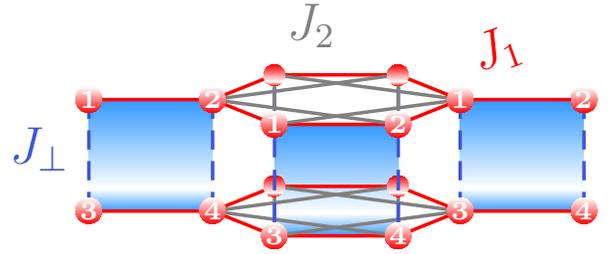}
\caption{(Color online)  Schematic representation of the relevant couplings interactions in \bimno.
Colored areas correspond to the unit cells. The sites in each unit cell are labeled from 1 to 4.}
\label{fig:layers}
\end{figure}
\end{center}
  In Ref. \onlinecite{expnew2}, Matsuda {\it et. al.} have
 found experimental evidence that $J_{1}$, $J_{2}$ and $J_{\bot}$ are the dominant couplings and there is competition between them. As a result of this competition,
  a disordered ground state has been found.
This observation has raised the interest in the study of
magnetically disordered phases in honeycomb lattice
antiferromagnets. Most of the advances have been reached in the
single layer honeycomb
lattice\cite{Wang,Cabra_honeycomb_prb,Cabra_honeycomb_2,Zhang_PRB_2013,Mulder,Mosadeq,Albuquerque,Reuther,Farnell,Bishop_2012,Li_2012_honeyJ1-J2-J3,Bishop_2013,Clark,Mezzacapo,Ganesh_PRL_2013,Zhu_PRL_2013,Fisher_2013,Okumura,Ganesh_2011}
whereas less attention has been given to the unfrustrated bilayer
case\cite{Ganesh_2011,Ganesh_QMC,Oitmaa_2012}. Therefore, there are still
many open issues, especially for the frustrated bilayer case.

The aim of this paper is to study the zero temperature phase diagram
of the frustrated Heisenberg model on the bilayer honeycomb lattice.
The dependence on the interlayer coupling $J_{\bot}$ is investigated
for different values of the frustration $J_{2}$. We focus on the
$S=1/2$ case, where quantum fluctuations become more important. The
present study has several motivations. On the one hand, the phase
diagram corresponding to $S=1/2$ is a natural extension of the
recently presented phase diagram for the single layer honeycomb
lattice\cite{Zhang_PRB_2013}. On the other hand, the substitution of
Mn$^{4+}$ in \bimno\ by V$^{4+}$ may lead to the realization of the
$S=1/2$ Heisenberg model on the honeycomb lattice.

In this paper, using the Schwinger boson representation followed by
a mean field decoupling, the presence of both magnetically ordered
and disordered phases is investigated. We observe signatures of a
reentrant behavior in the melting of N\'eel phase.
The behavior of the local magnetization as a function of the
interlayer coupling $J_{\bot}$ gives a physical explanation to this
effect, since a small $J_{\bot}$ makes the system more magnetically
ordered. Another key finding of our work is that the interlayer
coupling may restore the lattice rotational symmetry within layers.
Furthermore, the linear spin wave theory (LSWT) is used to describe
the general behavior as a function of the spin $S$. To support the
mean-field results Lanczos technique in small systems is used,
complemented with series expansion based on the continuous unitary
transformation method to estimate the ground-state energy and the
triplet gap. Last but not least, a comparison with previous
Schwinger boson mean-field results for the $S=1/2$ Heisenberg
model\cite{Zhang_PRB_2013} on the single-layer case is discussed.

The outline of the paper is as follows: In Sec. II we introduce the
model. In Sec. III we apply the Schwinger boson mean-field approach
for $S=1/2$ case, complemented with Lanczos technique. In Sec. IV we
apply LSWT for general spin $S$. A comparison of the ground-state
energy and the triplet gap obtained by means of series expansion and
Lanczos technique is presented in Sec. V. We close with a discussion
and conclusions in Sec. VI

%---------------------------------------------------------------------

\section{Frustrated Bilayer Heisenberg model}
The Heisenberg model on the bilayer honeycomb lattice is described by
\ba
 \label{eq:Hspin_general}
 H &=&\sum_{\vec{r},\vec{r}',\alpha,\beta}
 J_{\alpha,\beta}(\vec{r},\vec{r}')
 \vec{\bf{S}}_{\alpha}(\vec{r})\cdot \vec{\bf{S}}_{\beta}(\vec{r}')
\ea
%
% %
% \begin{widetext}
% \small
% \ba
%  \label{eq:Hspin_general}
% \nonumber
%  H &=&\sum_{\vec{r}} \left\{
%  J_1 \left[
%  \vec{\bf{S}}_{1}(\vec{r})\cdot \vec{\bf{S}}_{2}(\vec{r})
% +\vec{\bf{S}}_{1}(\vec{r})\cdot \vec{\bf{S}}_{2}(\vec{r}+\vec{e}_{1})
% +\vec{\bf{S}}_{2}(\vec{r})\cdot \vec{\bf{S}}_{1}(\vec{r}+\vec{e}_{2})
% %
% + \vec{\bf{S}}_{3}(\vec{r})\cdot \vec{\bf{S}}_{4}(\vec{r})
% +\vec{\bf{S}}_{3}(\vec{r})\cdot \vec{\bf{S}}_{4}(\vec{r}+\vec{e}_{1})
% +\vec{\bf{S}}_{4}(\vec{r})\cdot \vec{\bf{S}}_{3}(\vec{r}+\vec{e}_{2})
% \right]\right.\\
% %
% &+&\left.
% J_{2}\sum_{\alpha=1}^{4}\left[
%  \vec{\bf{S}}_{\alpha}(\vec{r})\cdot \vec{\bf{S}}_{\alpha}(\vec{r}+\vec{e}_{1})
% +\vec{\bf{S}}_{\alpha}(\vec{r})\cdot \vec{\bf{S}}_{\alpha}(\vec{r}+\vec{e}_{2})
% +\vec{\bf{S}}_{\alpha}(\vec{r})\cdot \vec{\bf{S}}_{\alpha}(\vec{r}+\vec{e}_{1}+\vec{e}_{2})
% \right]
% %
% +J_{\bot}\left[
% \vec{\bf{S}}_{1}(\vec{r})\cdot \vec{\bf{S}}_{3}(\vec{r})
% +\vec{\bf{S}}_{2}(\vec{r})\cdot \vec{\bf{S}}_{4}(\vec{r})
% \right]
% \right\}
% %
% \ea
% \normalsize
% \end{widetext}
% %
%
where, $\vec{\bf{S}}_{\alpha}(\vec{r})$ is the spin operator on site
$\alpha$ corresponding to the unit cell $\vec{r}$. $\alpha$ takes
the values $\alpha=1,2,3,4$ corresponding to the four sites on each
unit cell as depicted in Fig. \ref{fig:layers}.
The coupling constants $J_{\alpha,\beta}(\vec{r},\vec{r}')$ on the bonds of the bilayer
lattice are depicted in Figure \ref{fig:layers}.
%
% The unitary vectors
% $\vec{e}_{1}=\frac{1}{2}(\sqrt{3},3)$ and
% $\vec{e}_{2}=\frac{1}{2}(\sqrt{3},-3)$ are the primitive vectors
% corresponding to the triangular lattice.

The classical model displays N\'eel order for $J_{2}/J_{1}<1/6$. The
interlayer coupling $J_{\bot}$ does not introduce frustration in the
system and then, at the classical level and $T=0$, does not affect
the classical N\'eel phase. In the quantum case the situation is
much subtle, N\'eel order is likely to melt giving rise to a
non-magnetic phase. In the following, except being explicitly
specified, we fix the energy scale by taking $J_{1}=1$ to simplify
the notation.
For large values of $J_{\bot}$ we expect the ground state to be an
interlayer valence bond crystal (IVBC) with corresponding spins from
both layers forming dimers. There are two different ways to destroy
the N\'eel order, by increasing the frustration on each layer or
increasing the coupling between layers. The destruction of the
N\'eel order in a single layer honeycomb lattice due to the
frustration introduced by means of the next-nearest neighbors
interactions has been studied by various approaches, including spin
wave theory\cite{Rastelli,Fouet,Mulder,Ganesh_2011}, a non-linear
$\sigma$-model approach\cite{Einarsson}, Schwinger boson mean-field
theory (SBMFT)\cite{Mattsson_PRB_1994,Wang,Zhang_PRB_2013}, bond
operator mean-field theory\cite{Mulder}, exact diagonalization
(ED)\cite{Fouet,Mosadeq,Albuquerque}, a variational Monte Carlo
(VMC) method\cite{Clark,Mezzacapo}, series expansion
(SE)\cite{Oitmaaj1j2j3}, the pseudofermion functional
renormalization group (PFFRG)\cite{Reuther}, the coupled cluster
method (CCM)\cite{Bishop_2012,Li_2012_honeyJ1-J2-J3,Bishop_2013} and
the density matrix renormalization group (DMRG)
method\cite{Ganesh_PRL_2013,Zhu_PRL_2013,Fisher_2013}.

{ For the single layer case the most accepted scenario is that
at a critical value of the frustrating coupling $J_{2}$ the N\'eel
order is destroyed giving rise to a magnetically disordered phase.
The different techniques listed above have yielded strong evidence
supporting the existence of an intermediate magnetically disordered
region where a spin gap opens and spin-spin correlations decay
exponentially
\cite{Zhang_PRB_2013,Albuquerque,Mezzacapo,Bishop_2012,Li_2012_honeyJ1-J2-J3}.
This disordered region comprises two kinds of magnetically
disordered phases distinguished by a rotational symmetry
breaking\cite{Zhang_PRB_2013}. In the range $0.2075\lesssim
J_{2}/J_{1}\lesssim 0.3732 $ there is a gaped spin liquid (GSL)
phase, where the ground state is magnetically disordered and
preserves all the lattice symmetries\cite{Zhang_PRB_2013}.

For larger values of $J_{2}$ the system presents a ground state that
breaks the lattice rotational symmetry, but preserves lattice
translational symmetry. This staggered dimer valence-bond crystal
(VBC), which is also called lattice nematic\cite{Mulder} was found
by using a variety of
techniques\cite{Mosadeq,Albuquerque,Zhang_PRB_2013,Mulder,Clark,Ganesh_PRL_2013,Zhu_PRL_2013}.
Finally for $J_{2}/J_1 \gtrsim 0.398$, the system enters into a
spiral phase\cite{Zhang_PRB_2013,Mulder}. These phases (except the
spiral one) are depicted in Figure \ref{fig:Phasediag_Jp-J2}-II.}

\begin{center}
\begin{figure}[t!]
\includegraphics[width=1.05\columnwidth]{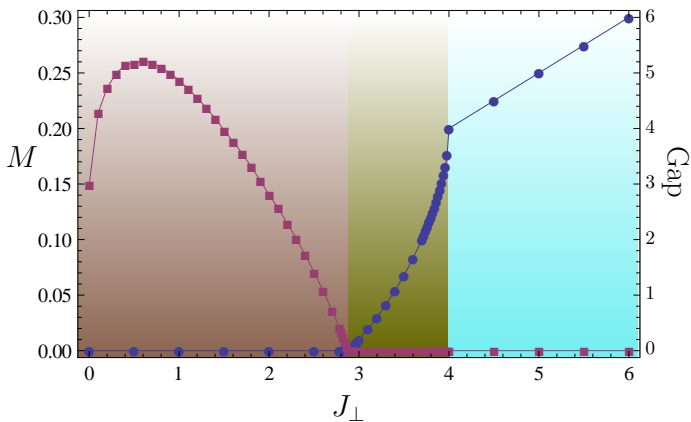}
\caption{(Color online)  The spin gap (blue circles) and sublattice
magnetization (red squares) obtained by SBMFT extrapolated to the
thermodynamic limit, corresponding to the dashed line in Fig.
\ref{fig:Phasediag_Jp-J2} ($J_{2}=0.1$). For $J_{\bot}> 4$ the gap
is proportional to $J_{\bot}$. Sublattice magnetization shows that
N\'eel order is enhanced by small interlayer coupling, reaching a
maximum at $J_{\bot} \sim 1/2$, after that it decreases until
disappearing at $J_{\bot}\sim 2.9$. {The brown shaded region
corresponds to the N\'eel phase. In the green and light-blue
regions, there is no evidence of any kind of magnetic order, and the
light-blue region presents a gap that depends linearly with
$J_{\bot}$.
%does not seem to be a genuine phase itself, but
%a signal of the breakdown of the mean field calculation near the transition.
}} \label{fig:gap_SBMFT}
\end{figure}
\end{center}

Since the nematic phase present in the single layer case breaks the
discrete rotational symmetry of the lattice, it is expected that, in
the bilayer case, by increasing the interlayer coupling the system
should undergo a transition from the nematic VBC to the IVBC phase,
restoring the $Z_{3}$ symmetry.

As shown in the following sections, we study the Hamiltonian
(\ref{eq:Hspin_general}) using a rotationally invariant
decomposition for the mean field parameters corresponding to a
Schwinger boson representation of the spin operators, which has
proven to be successful in incorporating quantum
fluctuations\cite{Zhang_PRB_2013,Cabra_honeycomb_prb,Cabra_honeycomb_2,Trumper1,Gazza,Trumper2,Coleman,Mezio,Feldner,Messio,Messio_2013}.
We complement this approach with ED, LSWT and SE.

\section{Schwinger Boson Mean-Field Approach and Exact Diagonalization}
\label{sec:sbmft}

In the Schwinger-boson representation, the Heisenberg interaction
can be written as a biquadratic form. The spin operators are
replaced by two species of bosons via the
relation\cite{Auerbach,Auerbach:1994,Auerbach:2011}
 \ba
 \vec{\mathbf{S}}_{\alpha}(\vec{r})=\frac{1}{2}\vec{\mathbf{b}}_{\alpha}^{\dag}(\vec{r})\cdot\vec{\sigma}\cdot\vec{\mathbf{b}}_{\alpha}(\vec{r}),
\ea
 where ${\vec{\bf b}_{\alpha}(\vec{r})^{\dagger }}\!=\!({\bf b}^{\dagger }_{\alpha,\uparrow }(\vec{r}),{\bf b}^{\dagger }_{\alpha,\downarrow }(\vec{r}))$
 is a bosonic spinor corresponding to the site $\alpha$ in the
 unit cell sitting at $\vec{r}$.  $\vec{\sigma}$ is the vector of
Pauli matrices, and there is a boson-number restriction
$\sum_\sigma \mathbf{b}^{\dag}_{\alpha,\sigma}(\vec{r})\mathbf{b}_{\alpha,\sigma}(\vec{r})\!=\!2S$ on each site.

In terms of boson operators we define the $SU(2)$ invariants
\ba
 \mathbf{A}_{\alpha \beta}(\vec{x},\vec{y})&=&\frac12 \sum_{\sigma} \sigma
\mathbf{b}_{\alpha,\sigma}(\vec{x})\mathbf{b}_{\beta,-\sigma}(\vec{y})\\
\mathbf{B}_{\alpha \beta}(\vec{x},\vec{y})&=&\frac12 \sum_{\sigma}
\mathbf{b}^{\dag}_{\alpha,\sigma}(\vec{x})\mathbf{b}_{\beta,\sigma}(\vec{y}).
\ea \normalsize
The operator $\mathbf{A}_{\alpha \beta}(\vec{x},\vec{y})$ creates a
spin singlet pair between sites $\alpha$ and  $\beta$ corresponding
to unit cells located at $\vec{x}$ and
 $\vec{y}$ respectively.
The operator $\mathbf{B}_{\alpha \beta}(\vec{x},\vec{y})$ creates a
ferromagnetic bond, which implies the intersite coherent hopping of
the Schwinger bosons.

\begin{center}
\begin{figure}[t!]
\includegraphics[width=0.98\columnwidth]{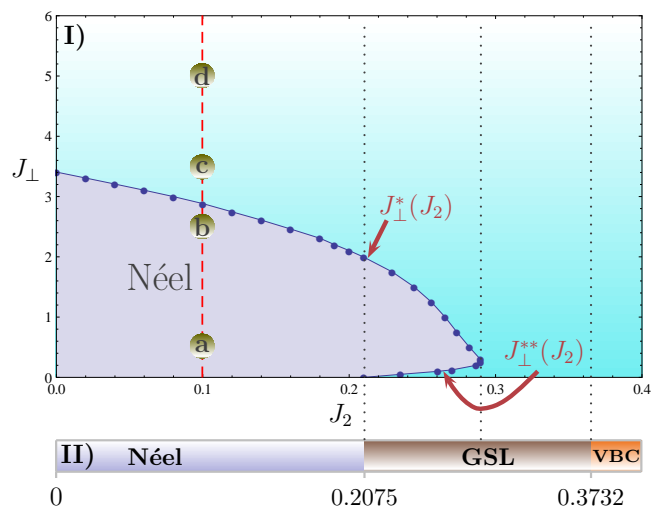}
\caption{ { (Color online) I) Phase diagram for $S=1/2$ in the
$J_{2}-J_{\bot}$ plane obtained by means of SBMFT. Gray region
correspond to the N\'eel phase whereas light-blue region corresponds
to magnetically disordered phases. Vertical dotted lines are used as
a reference showing the phases corresponding to the single-layer
case and the end of N\'eel phase re-entrace in $J_{2}-J_{\bot}$
plane. II) Phase diagram of the single layer case corresponding to
Ref. \onlinecite{Zhang_PRB_2013}.}} \label{fig:Phasediag_Jp-J2}
\end{figure}
\end{center}

In this representation, the rotational invariant spin-spin
interaction can be written as
\small
\ba
\nonumber
  \vec{\mathbf{S}}_{\alpha}(\vec{x})\cdot \vec{\mathbf{S}}_{\beta}(\vec{y})
  =:\mathbf{B}^{\dag}_{\alpha \beta}(\vec{x},\vec{y}) \mathbf{B}_{\alpha \beta}(\vec{x},\vec{y}):
   -\mathbf{A}^{\dag}_{\alpha \beta}(\vec{x},\vec{y}) \mathbf{A}_{\alpha \beta}(\vec{x},\vec{y})
\ea
\normalsize
 where $:\mathbf{O}:$ indicates the normal ordering of the operator $\mathbf{O}$.
One of the advantages of this rotational invariant decomposition is
that it enables to treat ferromagnetism and antiferromagnetism on
equal footing. This decomposition has been successfully used to
describe quantum disordered phases in two-dimensional frustrated
antiferromagnets\cite{Cabra_honeycomb_prb,Cabra_honeycomb_2,Zhang_PRB_2013,Trumper1,Trumper2,Mezio,Messio,Messio_2013}.
\begin{center}
\begin{figure*}[t!]
\includegraphics[width=1.8\columnwidth]{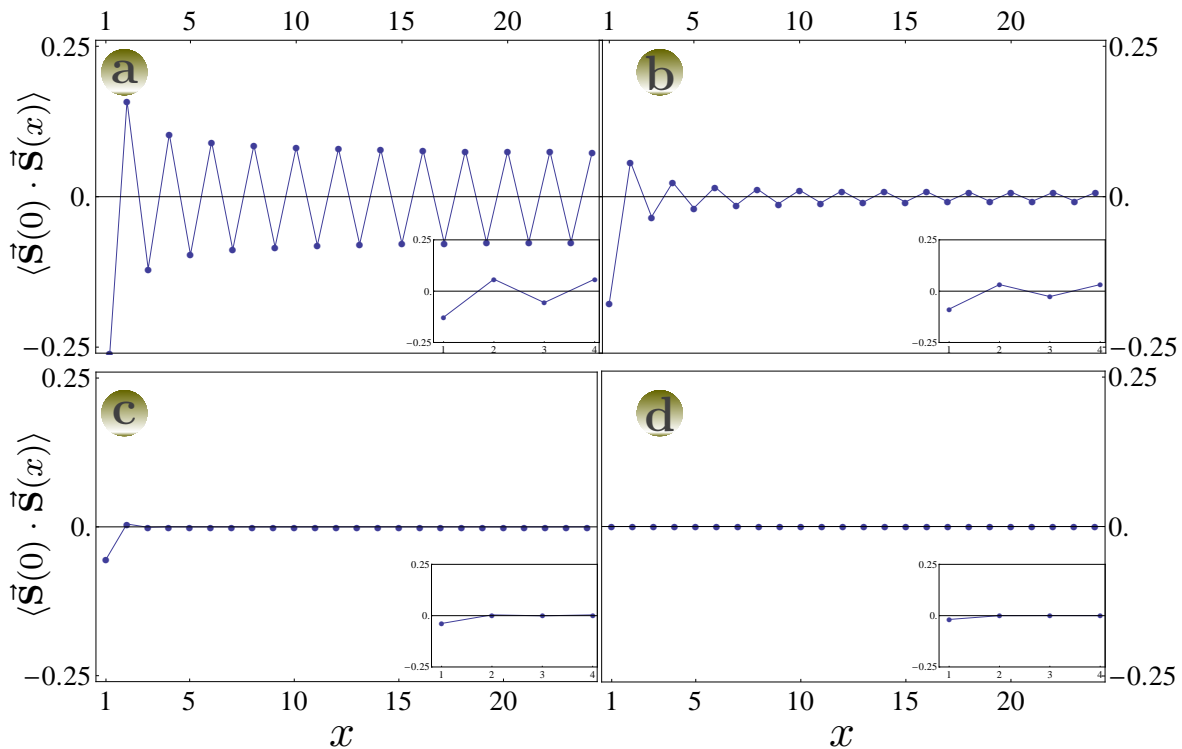}
\caption{(Color online)  { Spin-spin correlation
 between spins belonging to the same layer in the zig-zag direction} obtained by SBMFT
for a 10000 sites system. The labels a,b,c,d correspond to the
points showed in Fig. \ref{fig:Phasediag_Jp-J2} ($J_{\bot}=0.5, 2.5, 3.5, 5$). In the insets we
show the same correlations obtained by Lanczos diagonalization of a
24 sites system.} \label{fig:correlation_SS_lanczos}
\end{figure*}
\end{center}
In order to generate a mean field theory, we perform the Hartree-Fock decoupling
\ba
\nonumber
(\vec{\mathbf{S}}_{\alpha}(\vec{x})\cdot \vec{\mathbf{S}}_{\beta}(\vec{y}) )_{MF}&=&
 [B^{*}_{\alpha \beta}(\vec{x}-\vec{y}) \mathbf{B}_{\alpha \beta}(\vec{x},\vec{y})\\
 &-& A^{*}_{\alpha \beta}(\vec{x}-\vec{y}) \mathbf{A}_{\alpha \beta}(\vec{x},\vec{y})]\\\nonumber
&-& \langle (\vec{\mathbf{S}}_{\alpha}(\vec{x})\cdot \vec{\mathbf{S}}_{\beta}(\vec{y}) )_{MF}
\rangle
\ea
where the mean field parameters are given by
\ba
\label{eq:mfeqs}
A^{*}_{\alpha \beta}(\vec{x}-\vec{y})&=&\langle \mathbf{A}^{\dag}_{\alpha \beta}(\vec{x},\vec{y})\rangle,\\
\label{eq:mfeqs2}
B^{*}_{\alpha \beta}(\vec{x}-\vec{y})&=&\langle \mathbf{B}^{\dag}_{\alpha \beta}(\vec{x},\vec{y})\rangle ,
\ea
and the exchange at the mean field level is
\ba
\langle (\vec{\mathbf{S}}_{\alpha}(\vec{x})\cdot \vec{\mathbf{S}}_{\beta}(\vec{y}) )_{MF} \rangle
= |B_{\alpha \beta}(\vec{x}-\vec{y})|^{2}-|A_{\alpha \beta}(\vec{x}-\vec{y})|^{2}.\nn\\
\ea
The mean field equations (\ref{eq:mfeqs}) and (\ref{eq:mfeqs2}) must
be solved in a self-consistent way together with the following
constraint for the number of bosons in the system
\ba \label{eq:constraint} B_{\alpha
\alpha}(\vec{R}=\vec{0})=4N_{c}S, \ea
where $N_c$ is the total number of unit cells and $S$ is the spin
strength.
{ We use a self consistent procedure to find mean field
solutions that distinguish N\'eel phase from magnetically disordered
phases, and in particular possible phases without translational and
rotational symmetry breaking. Following the lines of Ref.
\onlinecite{Zhang_PRB_2013}, we work with two sites per unit cell,
which is  the smallest unit cell compatible with these kind of
solutions. }

Self consistent solutions in the bilayer honeycomb lattice involve
finding the roots of coupled nonlinear equations for the mean field
parameters and solving the constraints to determine the values of
the Lagrange multipliers $\lambda^{(\alpha)}$ which fix the number
of bosons in the system. We perform the calculations for large
systems and extrapolate the results to the thermodynamic limit.
Details of the self consistent calculation can be consulted in the
bibliography\cite{Cabra_honeycomb_prb,Zhang_PRB_2013}.
%

%
%
%%%%%%%%%%%%% begin bloque 1
%
 Using SBMFT we study some features of the phase diagram in the $J_{2}-J_{\bot}$ plane.
The line $J_{\bot}=0$ corresponds to the phase diagram for the
single layer honeycomb lattice. A description of the phases
presented in the single layer phase diagram was obtained recently
using the same rotational invariant mean-field
decoupling\cite{Zhang_PRB_2013}.
One of the advantages of the SBMFT is that allows to study large systems and perform the extrapolation to the thermodynamic limit. In particular,
this is useful to determine whether the system remains gapless or not.
To obtain the phase boundary between the magnetically ordered and disordered phases we use the extrapolation of the gap in the boson spectrum.
In the gapless region the excitation spectrum is zero at $\vec{k}=\vec{0}$, where the boson condensation occurs, this is characteristic of the N\'eel ordered phase.
On the other hand, in the gapped region, the absence of Bose condensation indicates that the
ground state is magnetically disordered.
In Figure \ref{fig:gap_SBMFT}
 the extrapolation of the spin gap corresponding to $J_{2}=0.1$ (dashed line in
Fig.~\ref{fig:Phasediag_Jp-J2}) is presented as a function of the interlayer coupling $J_{\bot}$. For small
 values of the interlayer coupling the system remains gapless. As we increase $J_{\bot}$ the gap opens at a given value
 $J^{*}_{\bot}(J_{2})$. Increasing more the interlayer coupling, the gap becomes a linear function of
 $J_{\bot}$.
At the value $J^{*}_{\bot}(J_{2})$ the N\'eel order is destroyed
leading to  the IVBC ground state.  As is known,
mean-field techniques are not the most convenient methods to study
the properties of a system near a phase transition,
so it may be difficult to determine quantitatively the transition
between N\'eel and disordered phases using only SBMFT. For this
reason, in our case, we would tend to conclude that the abrupt
change of behavior in the gap $\Delta$ at $J_{\bot}=4$, Fig. ~\ref{fig:gap_SBMFT} does not
indicate a phase transition, but could be an indication of the breakdown of the mean field
calculation. Actually physical quantities, as magnetization and correlations, calculated in green and light-blue shadowed regions
of Fig. ~\ref{fig:gap_SBMFT} do not show qualitative differences.

In Fig.~\ref{fig:Phasediag_Jp-J2} we show the phase diagram in the
$J_{2}-J_{\bot}$ plane corresponding to $S=1/2$. For $J_{\bot} \gg
J_2$ one can expect a IVBC ground state adiabatically  connected
with the limit of decoupled dimers, i.e. two singlets per unit cell,
between spins $1(2)$ and $3(4)$ (see Fig. \ref{fig:layers}).
In this limit the ground state energy per dimer is
$E_{IVBC}=-\frac{3}{4}J_{\bot}$, with an energy gap
$\Delta=J_{\bot}$ to triplet magnetic excitations.

In order to support the analytical results of the mean-field
approach, we have also performed Lanczos ED calculations on finite
systems with 24
 spins and periodic boundary conditions for $S=1/2$.
 The bilayer structure of the lattice makes particularly difficult to study small systems because
 there
 are four sites per unit cell. In particular, correlation functions between spins belonging to
 the same layer can be studied only for a few neighbors.
 Fig.~\ref{fig:correlation_SS_lanczos} shows the spin-spin correlation
 between spins belonging to the same layer in the zig-zag direction obtained by SBMFT corresponding to
 the points (a), (b), (c) and (d) of Fig.~\ref{fig:Phasediag_Jp-J2} for a 10000 sites system. The insets correspond to the results
 obtained for the same points with Lanczos technique on a 24 sites system.
 Although correlations are  calculated only for a few sites with Lanczos, the absence of
 antiferromagnetic order in the insets of Figures 4.c and 4.d is clear.
 This is consistent with the SBMFT results corresponding to the main Figures 4.c and 4.d.

\begin{center}
\begin{figure}[t!]
\includegraphics[width=0.9\columnwidth]{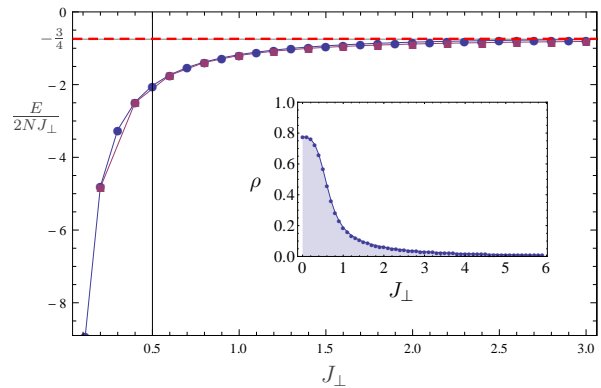}
\caption{(Color online) Ground state energy per dimer (in units of
$J_{\bot}$) as a function of $J_{\bot}$ for $J_{2}=0.18$ obtained by
means of SBMFT extrapolated to the thermodynamic limit (blue
circles) and Lanczos for a 24 sites system (red squares). Horizontal
dashed red line indicates to $\frac{E}{2NJ_{\bot}}=-3/4$
corresponding to the  decoupled dimer product state. Inset:
$Z_3$ directional symmetry-breaking order parameter $\rho$ as a
function of $J_{\bot}$ corresponding to the line $J_{2}=0.38$.}
\label{fig:E_overJ3}
\end{figure}
\end{center}

In Fig.~\ref{fig:E_overJ3} we show the energy per dimer in units of
$J_{\bot}$ calculated with SBMFT (blue circles)
 and Lanczos for a system with 24 sites (red squares) for $J_{2}=0.18$.
 As it can be observed the energy per dimer gets very close to the value corresponding to a dimer product state
 $\displaystyle{\frac{E}{2N J_{\bot}}} =-3/4$, as $J_{\bot}$ is increased.

As showed in Fig. \ref{fig:Phasediag_Jp-J2}, in
the region $0.2075\lesssim J_{2}\lesssim 0.289$ there is a reentrant
effect. In this range, N\'eel phase separates from $J_2$ axis,
leaving a tiny space for a magnetically disordered phase.
In this way, N\'eel phase is here not only limited by some value $J^{*}_{\bot}(J_2)$
 from above, but also by a second value $J^{**}_{\bot}(J_2)$
(See Figure \ref{fig:Phasediag_Jp-J2}) from below. In
Fig.~\ref{fig:gap_SBMFT}, we show the sublattice
magnetization\cite{Zhang_PRB_2013,Trumper1,Gazza} along the line
$J_{2}=0.1$. It is clear that a small bilayer coupling enhances the
antiferromagnetic long range order, which is the reason of the
reentrant effect.

On the other hand, in the range $0.3732 \lesssim J_{2} \lesssim
0.398 $ ($J_{\bot}=0$), there is evidence of the existence of an
on-layer nematic VBC phase\cite{Zhang_PRB_2013} {(see Figure \ref{fig:Phasediag_Jp-J2}-II)}.
In this VBC phase SU(2) spin rotational and lattice translational
symmetries are preserved. But $Z_3$ symmetry, corresponding to
$2\pi/3$ rotations around an axis perpendicular to the plane and
passing through a site, is broken.
By increasing the interlayer coupling $J_{\bot}$ the system moves to the IVBC where the $Z_{3}$
symmetry is recovered.
In order to observe this symmetry restoring we introduce the
$Z_3$ directional symmetry-breaking order parameter
$\rho$\cite{Okumura,Mulder}

\ba
\nonumber
\rho&=&\frac43 |(\langle \vec{\mathbf{S}}_{1}(\vec{r}) \cdot \vec{\mathbf{S}}_{2}(\vec{r}) \rangle
+e^{i2\pi/3} \langle \vec{\mathbf{S}}_{1}(\vec{r}) \cdot \vec{\mathbf{S}}_{2}(\vec{r}+\vec{e}_{1}) \rangle \\
&+& e^{i4\pi/3} \langle \vec{\mathbf{S}}_{1}(\vec{r}) \cdot \vec{\mathbf{S}}_{2}(\vec{r}-\vec{e}_{2}) \rangle )|
\ea
This order parameter is zero when the bond energies $\kappa_{i}=
\langle \vec{\mathbf{S}}_{1}(\vec{r}) \cdot
\vec{\mathbf{S}}_{2}(\vec{r}+\vec{e}_{i}) \rangle $ are equal, being
$i=0,1,2$ where $\vec{e}_{0}=\vec{0}$, $\vec{e}_{1}=\frac{1}{2}(\sqrt{3},3)$ and
$\vec{e}_{2}=\frac{1}{2}(\sqrt{3},-3)$. The parameter is chosen to be
$\rho=1$ when only one of the bond energies is nonzero.
In the case $J_{\bot}=0$, this parameter is nonzero in the region
$0.3732 \lesssim J_{2} \lesssim 0.398 $. For small interlayer
coupling, the bond energies satisfy
$\kappa_{i}\neq\kappa_{j}=\kappa_{k}$. Therefore, the system is still
in the nematic VBC phase.
 But increasing further the inter-layer coupling the order parameter tends to zero
continuously as shown in the inset of Figure \ref{fig:E_overJ3} and for large $J_{\bot}$
the system enters in the IVBC.
Finally, in the region $0.289 \lesssim J_2 /J_1 \lesssim 0.3732$ the
ground state preserves SU(2), lattice translational and  $Z_3$
symmetries and the spin-spin correlations are short ranged. This
agrees with the evidence of a spin liquid phase in the phase diagram
corresponding to $J_{\bot}=0$\cite{Clark,Mezzacapo,Zhang_PRB_2013}.

\section{Linear spin wave Theory}

In this section we use a linear spin wave approach to study the stability of N\'eel order as a function of the spin strength.
The classical spin state corresponding to the energy minimum of the Hamiltonian (\ref{eq:Hspin_general})
for $J_{2}<\frac16 J_{1} $ is given by an anti-parallel  (N\'eel) configuration.
Incorporating quantum fluctuation to the classical ground state is
likely to lead to the melting of N\'eel order.
\begin{center}
\begin{figure}[t!]
\includegraphics[width=0.98\columnwidth]{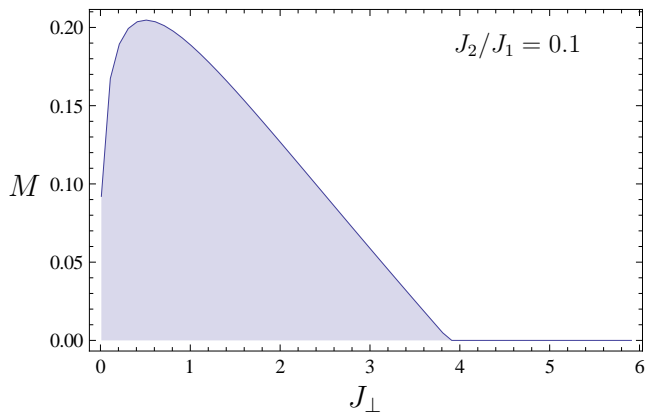}
\caption{(Color online) Staggered magnetization vs. $J_{\bot}$ obtained by means of linear spin wave
approximation. For small values of the intralayer coupling the N\'eel order is enhanced,
in agreement with SBMFT results. }
\label{fig:mag_SW}
\end{figure}
\end{center}

For the spin wave implementation it is convenient to define new spin
operators by rotating in $\pi$ the spins  belonging to sublattices
2 and 3 (See Fig. \ref{fig:layers}) about the  $x$-axis. After the
rotation we have
\ba
\tilde{S}^{x}_{\alpha}(\vec{r})&=&S^{x}_{\alpha}(\vec{r})\\
\tilde{S}^{y}_{\alpha}(\vec{r})&=&-S^{y}_{\alpha}(\vec{r})\\
\tilde{S}^{z}_{\alpha}(\vec{r})&=&-S^{z}_{\alpha}(\vec{r}),
\ea
for spin operators belonging to sublattices 2 or 3, while
$\tilde{\vec{S}}_{\alpha}(\vec{r})=\vec{S}_{\alpha}(\vec{r})$ for
sublattices 1 and 4. Thereby, the classical ground state have all
spins pointing towards the new $S^{z}$ axis.

\begin{center}
\begin{figure}[t!]
\includegraphics[width=0.98\columnwidth]{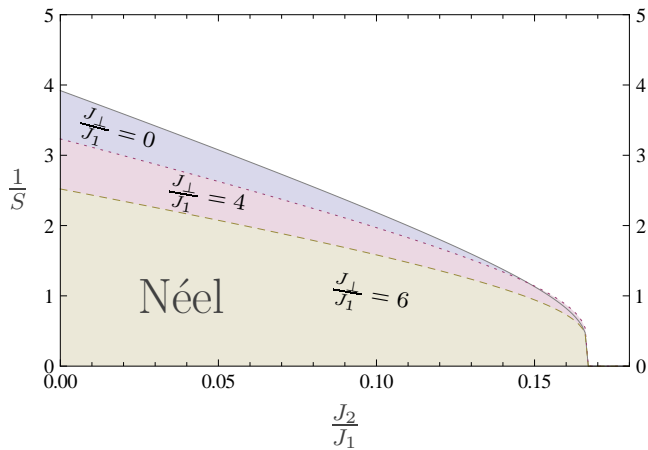}
\caption{(Color online)  Phase diagram in the $1/S$-$J_{2}$ plane for different values of  $J_{\bot}$
obtained by means of LSWT. }
\label{fig:ScvsJ2}
\end{figure}
\end{center}

In order to study spin-wave fluctuations, we write the spin operators in
terms of Holstein-Primakoff bosons as
follows
\ba
\tilde{S}^{+}_{\alpha}(\vec{r})&=&\sqrt{2S}a_{\alpha}(\vec{r})\\
\tilde{S}^{-}_{\alpha}(\vec{r})&=&\sqrt{2S}a^{\dagger}_{\alpha}(\vec{r})\\
\tilde{S}^{z}_{\alpha}(\vec{r})&=&S-n_{\alpha}(\vec{r}).
\ea
The Hamiltonian can be written in terms of these boson operators as
\ba
H=E_{0}+H_{SW},
\ea
with
\ba
E_{0}&=&2NS^{2}(6J_{2}-3J_{1}-J_{\bot}) \\\nonumber
H_{SW}&=&2NS(6J_{2}-3J_{1}-J_{\bot})\\
 &+&\int d^{2}\vec{k}\;
 \vec{\mathbf{a}}^{\dagger}(\vec{k})\mathcal{M}(\vec{k})\vec{\mathbf{a}}(\vec{k}),
\ea
where $\vec{\mathbf{a}}(\vec{k})$ is a vector of bosonic operators.
% %
% \ba
% \vec{\mathbf{a}}(\vec{k})=\left(
% \begin{tabular}{c}
% $ \mathbf{a}_{1}(\vec{k})$\\
% $\mathbf{a}_{2}(\vec{k})$\\
% $\mathbf{a}_{3}(\vec{k})$\\
% $\mathbf{a}_{4}(\vec{k})$\\
% $\mathbf{a}^{\dagger}_{1}(-\vec{k})$\\
% $\mathbf{a}^{\dagger}_{2}(-\vec{k})$\\
% $\mathbf{a}^{\dagger}_{3}(-\vec{k})$\\
% $\mathbf{a}^{\dagger}_{4}(-\vec{k})$
% \end{tabular}
% \right),
% \ea
% %
%
% %
% \ba
% \vec{\mathbf{a}}^{\dagger}(\vec{k})=\left(
%  \mathbf{a}_{1}(\vec{k}), \mathbf{a}_{2}(\vec{k}),
% \mathbf{a}_{3}(\vec{k}), \mathbf{a}_{4}(\vec{k}),
% \mathbf{a}^{\dagger}_{1}(-\vec{k}), \mathbf{a}^{\dagger}_{2}(-\vec{k}),
% \mathbf{a}^{\dagger}_{3}(-\vec{k}), \mathbf{a}^{\dagger}_{4}(-\vec{k})
% \right),
% \ea
% %
%and
%
\ba
\mathcal{M}(\vec{k})=\left(
\begin{tabular}{cc}
 $\Gamma$ & $\Omega$ \\
 $\Omega$ & $\Gamma$
\end{tabular}
\right)
\ea
with
\ba
\Gamma=\left(
\begin{tabular}{cccc}
$\gamma_{2}(\vec{k})$ & $0$ & $0$ & $0$\\
 $0$ & $\gamma_{2}(\vec{k})$ & $0$ & $0$ \\
 $0$ & $0$ & $\gamma_{2}(\vec{k})$ & $0$\\
 $0$ & $0$ & $0$ & $\gamma_{2}(\vec{k})$
\end{tabular}
\right)\\
\Omega=\left(
\begin{tabular}{cccc}
$0$ & $\gamma_{1}(-\vec{k})$ & $\gamma_{\bot}$ & $0$\\
 $\gamma_{1}(\vec{k})$ & $0$ & $0$ & $\gamma_{\bot}$ \\
 $\gamma_{\bot}$ & $0$ & $0$ & $\gamma_{1}(-\vec{k})$\\
 $0$ & $\gamma_{\bot}$ & $\gamma_{1}(\vec{k})$ & $0$
\end{tabular}
\right)
\ea
%
%
% \begin{widetext}
% %
% \ba
% \mathcal{M}(\vec{k})=\left(
% \begin{tabular}{cccccccc}
% $\gamma_{2}(\vec{k})$ & $0$ & $0$ & $0$ & $0$ & $\gamma_{1}(-\vec{k})$ & $\gamma_{\bot}$ & $0$  \\
% $0$ & $\gamma_{2}(\vec{k})$ & $0$ & $0$ & $\gamma_{1}(\vec{k})$ & $0$ & $0$ & $\gamma_{\bot}$  \\
% $0$ & $0$ & $\gamma_{2}(\vec{k})$ & $0$ & $\gamma_{\bot}$ & $0$ & $0$ & $\gamma_{1}(-\vec{k})$  \\
% $0$ & $0$ & $0$ & $\gamma_{2}(\vec{k})$ & $0$ & $\gamma_{\bot}$ & $\gamma_{1}(\vec{k})$ & $0$  \\
% $0$ & $\gamma_{1}(-\vec{k})$ & $\gamma_{\bot}$ & $0$ & $\gamma_{2}(\vec{k})$ & $0$ & $0$ & $0$  \\
% $\gamma_{1}(\vec{k})$ & $0$ & $0$ & $\gamma_{\bot}$ & $0$ & $\gamma_{2}(\vec{k})$ & $0$ & $0$  \\
% $\gamma_{\bot}$ & $0$ & $0$ & $\gamma_{1}(-\vec{k})$ & $0$ & $0$ & $\gamma_{2}(\vec{k})$ & $0$  \\
% $0$ & $\gamma_{\bot}$ & $\gamma_{1}(\vec{k})$ & $0$ & $0$ & $0$ & $0$ & $\gamma_{2}(\vec{k})$  \\
% \end{tabular}
% \right),
% \ea
% %
% \end{widetext}
%
%
being functions $\gamma_{1}$, $\gamma_{2}$ and $\gamma_{\bot}$ given by
\ba
\gamma_{1}(\vec{k})&=&\frac{1}{2}J_{1}S (1+e^{i\vec{k}\cdot\vec{e}_{1}}+e^{-i\vec{k}\cdot\vec{e}_{2}})\\\nonumber
\gamma_{2}(\vec{k})&=&J_{2}S (\cos(\vec{k}\cdot\vec{e}_{1})+\cos(\vec{k}\cdot\vec{e}_{2})
+\cos(\vec{k}\cdot(\vec{e}_{1}+\vec{e}_{2})))\\
&+& \frac{3}{2}J_{1}S-3J_{2}S+\frac{1}{2}J_{\bot}S\\
\gamma_{\bot}&=&\frac{1}{2}J_{\bot}S.
\ea
Then we use a Bogoliubov transformation to diagonalize the Hamiltonian $H_{SW}$
and obtain the following eigenvalues
\ba
\varepsilon^{\pm}_{\beta}(\vec{k})=\sqrt{(\gamma_{2}(\vec{k}))^{2} - (\gamma_{\bot}\pm|\gamma_{1}(\vec{k})|)^{2}},
\ea
where $\beta=1,2$ is the layer index. The staggered magnetization can be calculated in the linear approximation as follows
\begin{eqnarray}
\label{eq:magnetization1}
M &=&\frac{1}{4N}\sum_{\vec{r},\alpha }\tilde{S}_{\alpha }^{z}\left( \vec{r}%
\right)  \nonumber\\
&=&S-\frac{1}{4N}\sum_{\vec{r},\alpha }\mathbf{a}_{\alpha }^{\dagger }(\vec{r%
})\mathbf{a}_{\alpha }(\vec{r}).
\end{eqnarray}
%
% \ca{On one hand as $J_{2}$ is increased from zero, fluctuations around the N\'eel state increase
% due to frustration, and can destroy the N\'eel order.}
On the one hand, fluctuations around the N\'eel state increase
with the frustration $J_{2}$, and can destroy the N\'eel order.
On the other hand, as the value of S is lowered, quantum fluctuations become more important and we can expect the N\'eel state to melt for a given value of S.
The correction to the classical boundary for the N\'eel state can be estimated by finding the frustration $J_{2}$ at which the sublattice magnetization  $M$
given by Eq. (\ref{eq:magnetization1}) vanishes.
In Fig. \ref{fig:mag_SW} we present the sublattice magnetization $M$
as a function of the interlayer coupling ($J_{\bot}$) corresponding
to $J_{2}=0.1$ (dashed line on Figure \ref{fig:Phasediag_Jp-J2}).
Notice that, for small values of the interlayer coupling, the magnetization is an increasing
function of $J_{\bot}$,  i.e, the antiferromagnetic order is enhanced.
 But increasing more the value of $J_{\bot}$ the sublattice magnetization
 is reduced and vanishes for large values of $J_{\bot}$.
 This behavior is in agreement with the SBMFT results and the reentrant
 effect observed in Fig.~\ref{fig:Phasediag_Jp-J2}. In Fig. \ref{fig:ScvsJ2}, we present the melting curves in the $1/S$-$J_{2}$ plane for different values of $J_{\bot}$.
\begin{center}
\begin{figure*}
\includegraphics[width=1.5\columnwidth]{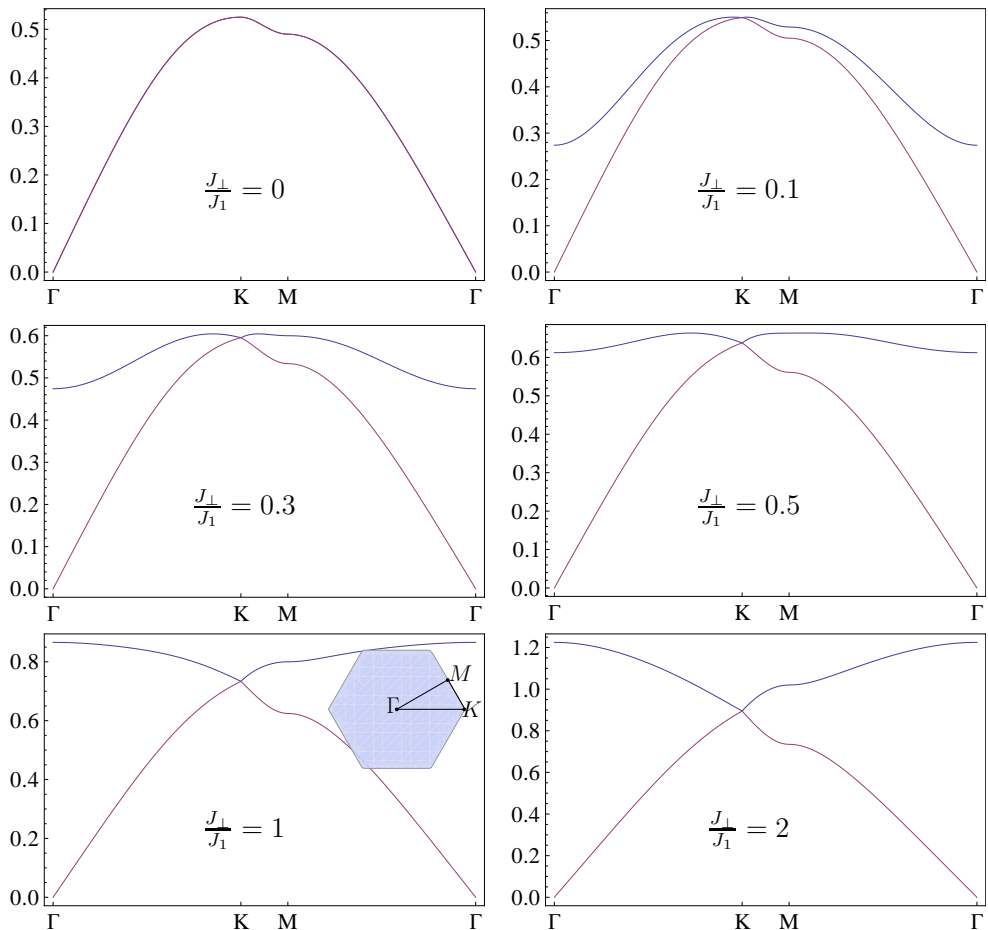}
\caption{(Color online)  Dispersion of magnon modes corresponding to $J_{2}/J_{1}=0.1$
along the path depicted in the inset, obtained by means of LSWT.}
\label{fig:dispersion}
\end{figure*}
\end{center}
The case
corresponding to $J_{\bot}=0$ agrees with LSWT results presented in
Ref. \onlinecite{Mattsson_PRB_1994} (Fig. 5) for the single layer
$J_{1}-J_{2}$ Heisenberg model. For large values of $J_{\bot}$ the
stability region for the N\'eel state is reduced. Notice that in
References \onlinecite{Cabra_honeycomb_prb} and \onlinecite{Mattsson_PRB_1994}
% \ca{there is a discrepancy between the boundary of the N\'eel state determined
% by LSWT and Schwinger bosons mean field.}
there is a discrepancy between LSWT and Schwinger bosons mean field
determinations of the N\'eel state boundary. This difference could
be reduced by means of higher order $1/S$ corrections to the LSWT,
which are beyond the scope of the present work.
Finally, in Fig.~\ref{fig:dispersion} we show the dispersion of
magnon modes along the path depicted in the inset. It is clear that
as we increase the interlayer coupling, two of the four magnon modes
acquires a nonzero gap at the $\Gamma$ point.

\section{Comparison between Series Expansion and Exact Diagonalization}

As a complement to our analysis, we have performed series expansion
(SE) calculations, starting from the limit of isolated dimers
connecting spins from both layers via $J_{\bot}$. This allows us to
assess Lanczos results by comparison with other non-mean-field
technique. To this end, we have decomposed the Hamiltonian
Eq.(\ref{eq:Hspin_general}) into $H=H_{0}(J_{\bot})+V(J_{1},J_{2})$,
where $H_{0}(J_{\bot})$ represents decoupled dimers and
$V(J_{1},J_{2})$ is the part of Hamiltonian that connects dimers by
means of $J_{1},J_{2}$ couplings.

Since each dimer has two energy levels (singlet and triplet), the
spectrum of $H_{0}(J_{\bot})$ is equidistant, allowing to sort the
levels structure of $H_{0}$ in a block-diagonal form, where each
block is labeled by an energy quantum-number Q. Ground state
(\emph{vacuum}) is in Q=0 sector, i.e., all dimers are in the
singlet state. Q=1 sector is composed by states obtained by creating
(from vacuum state) one-elementary triplet excitation
(\emph{particle}) on a given dimer, whereas that $Q\geq2$ is of
multiparticle nature.

Perturbation $V(J_{1},J_{2})$ does not conserve the block-diagonal form of $H_{0}(J_{\bot})$,
i.e., it mixes different Q-sectors. However, for this type of Hamiltonian, it can be shown \cite{Knetter2000a} that it is possible to recover the block-diagonal form
by means of continuous unitary transformations, using the flow equation method of Wegner
\cite{Wegner1994a}. It essentially consists in transforming $H$ onto an effective Hamiltonian $H_{\mathrm{eff}}$ which is block-diagonal in the quantum number $Q$. This transformation
can be achieved order by order in a perturbative series in powers of $J_{1,2}$, leading to
$H_{\mathrm{eff}}=H_{0}(J_{\bot})+\sum_{n,0\leq m\leq n}{J_{1}^{n-m}J_{2}^{m}C_{n,m}}$,
where $C_{n,m}$ are weighted products of Q-conserving terms in $V(J_{1},J_{2})$,
determined by recursive differential equations, see Ref. \onlinecite{Knetter2000a} for details.

$Q$-number conservation allows the evaluation of several observables
directly from $H_{\mathrm{eff}}$ in terms of a SE in $J_{1,2}$. For
the present model we have performed $O(5)$ and $O(4)$ SE in
$J_{1,2}$ for ground state energy ($Q=0$) and for triplet dispersion
($Q=1$), respectively. Explicit expressions are too long to be
printed explicitly, in particular triplet dispersion. Upon request
they will be made available electronically. Regarding technical
details about the calculation we refer to Ref.
\onlinecite{Arlego2011}.

\begin{center}
\begin{figure}
\includegraphics[width=0.9\columnwidth]{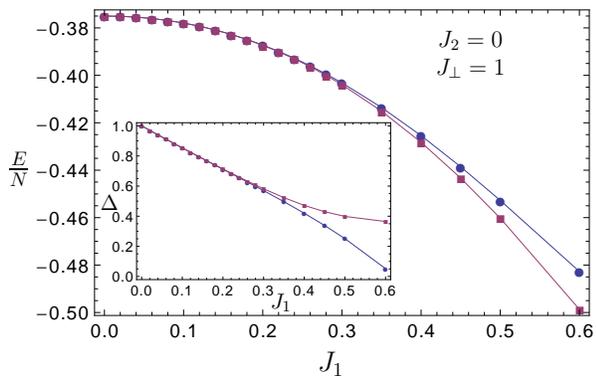}
\caption{(Color online) Ground state energy per site as a function of $J_{1}$
obtained by Lanczos (ED) on a $24$ sites system (red squares) and $O(5)$ Series
Expansion (SE) (blue circles).
Inset:  triplet gap (same set of parameters as main panel)
ED (red squares) and $O(4)$ SE (blue circles).}
\label{fig:SE}
\end{figure}
\end{center}

To illustrate the type of results obtained, in Fig. \ref{fig:SE} we
show the ground state energy per site as a function of $J_1$ and
$J_2=0$, obtained by O(5) SE (blue circles) and ED on a finite
system of 24 sites (red squares). As it can be observed, both
techniques predict an energy decreasing with the coupling of
interlayer-dimers via $J_1$. Furthermore, there is an excellent
quantitative agreement between both methods, up to $J_1 \simeq
0.25$. Beyond this value, the difference between the two approaches
becomes increasingly noticeable, being attributable to finite size
effects of ED and the order achieved in the SE. When the frustration
$J_2$ is incorporated, the agreement is not as good as in the
unfrustrated case.
This might be due to the stronger effect that the frustration induces on finite size effects in Lanczos calculation.\\
On the other hand, triplet gap is shown in the inset of Fig.
\ref{fig:SE} for the same set of parameters as the ground state
energy. Here we also observe that both techniques predict a tendency
to a closure of the gap, when $J_1$ is turned on. While in this case
we have achieved a $O(4)$ SE, we see that the range of agreement
between ED and SE is practically the same as before, being as well
reduced when the frustration is included. Overall, our calculations
shows that ED and SE share a range parameters where both predict the
same behavior. A more detailed analysis in search of transitions,
involving gap closure or level-crossings from SE point of view, is
beyond the scope of present work.

%\section{Results}

\section{Discussion and conclusions}

We have studied the phase diagram corresponding to a frustrated
Heisenberg model on the bilayer honeycomb lattice, by means of
Schwinger bosons mean field theory, complemented with exact
diagonalization, linear spin-wave theory and series expansion.

By analyzing the sublattice magnetization and the spin gap by SBMFT,
we have described the behavior of the quantum phases as the
interlayer coupling is increased. The absence of N\'eel order for
large values of the interlayer coupling has also been observed by
spin-spin correlation calculations, where SBMFT and ED techniques
predict the same qualitative behavior.

In particular, in the small frustration region {
($J_{2}/J_{1}\lesssim 0.2075$)} the system is N\'eel ordered for
$J_{\bot}=0$, but increasing the interlayer coupling up to a value
($J^{*}_{\bot}$) the N\'eel order is destroyed and the system enters
in a non magnetic phase.
The spin-spin correlations are
consistent with the destruction of the N\'eel order, given place, for large values of the
interlayer coupling, to a phase with short range spin-spin correlations and a finite spin gap.

In the region ($0.2075\lesssim J_{2}/J_{1} \lesssim 0.289$), the
phase diagram shows signatures of a reentrant behavior. At $J_{\bot}=0$ the system
does not present magnetic order, but increasing the interlayer
coupling up to a finite (and small) value $J^{**}(J_{2})$, the
system becomes N\'eel ordered. Increasing even more the interlayer
coupling, the N\'eel order is destroyed at a given value
$J^{*}(J_2)$. The behavior of the sublattice magnetization as the
function of $J_{\bot}$ also support the existence of the reentrant
behavior.

For values of the interlayer coupling between { ($ 0.289\lesssim
J_{2}/J_{1} \lesssim 0.3732 $)} the N\'eel order is absent at
$J_{\bot}=0$ and the system presents a nonzero spin gap, whereas in
the region { ($0.3732\lesssim J_{2}/J_{1} \lesssim 0.398$)}
 each layer presents a nematic disordered phase\cite{Zhang_PRB_2013}. In both cases
 increasing the value of $J_{\bot}$ the system goes to an
interlayer valence bond crystal with a spin gap that is proportional to the interlayer coupling.

In all the range of values $0 < J_{2}/J_{1} < 0.398 $, for
$J_{\bot}/J_{1}>4$ the system presents signatures of an
interlayer-valence bond crystal (IVBC) phase that evolves
adiabatically from the limit of decoupled interlayer-dimers. This is
corroborated by series expansion calculations starting explicitly
from the limit of isolated interlayer dimers.

The precise determination of transitions lines between different
quantum phases present in the model is not a simple task. Among
these issues, the important question about how are the nematic VBC
and IVBC phases connected still remains open. The mean field
character of SBMFT method does not allow us to draw a definite
conclusion about the real nature of the transition. From the
viewpoint of series expansion (SE), it is possible to analyze the
adiabatic evolution of the nematic VBC phase, starting appropriately
from isolated dimers on each plane. Thus, the possibility and type
of transitions between nematic VBC and IVBC phases could be
estimated by analyzing level crossings and gap closures between this
SE and the dimer SE obtained in Section V for IVBC phase. While this
analysis goes beyond the scope of this work, it clearly deserves
more investigation. We postpone the detailed study of these
transitions for future work, as we have focused on the general
characteristics of each region of the phase diagram.

\section*{ACKNOWLEDGMENTS}

We are especially grateful to Hirokazu Tsunetsugu for fruitful
discussions. H.Zhang is supported by the Japanese Government
Scholarship from MEXT of Japan. C. A. Lamas and M. Arlego are
partially supported by CONICET (PIP 1691) and ANPCyT (PICT 1426).

%%%%%%%%%%%%%%%%%%%%%%%%%%%%%%%%%%%%%%%%%%%%%%%%%%%%%%%%%%%%%%%%%%%%%%%%%%%%%%%%


\begin{thebibliography}{}



%1
\bibitem{Anderson} P.~W.~Anderson,
Mater.\ Res.\ Bull.\ {\bf 8}, 153 (1973); P.~Fazekas and
P.~W.~Anderson, Phil.\ Mag.\ {\bf 30}, 423 (1974); P.~W.~Anderson,
Science {\bf 235}, 1196 (1987).

%2
\bibitem{Balents_Nature} L.~Balents, Nature {\bf 464}, 199 (2010).

%3
\bibitem{ML_2005} G.\ Misguich and C.\ Lhuillier, in {\it Frustrated Spin Systems}, edited by H.~T.~Diep, (World Scientific, Singapore, 2005).

\bibitem{Oshikawa_2013} C. A. Lamas, A. Ralko, M. Oshikawa, D. Poilblanc, and P. Pujol,Phys.\ Rev.\ B { \bf 87}, 104512 (2013)

%4
\bibitem{LM_2011} C.\ Lhuillier and G.\ Misguich, in {\it Introduction to Frustrated Magnetism}, Eds. C.\ Lacroix, P.\ Mendels, and F.\ Mila, (Springer-Verlag, Berlin Heidelberg, 2011).

%5
\bibitem{Wang} F.\ Wang, Phys.\ Rev.\ B {\bf 82}, 024419 (2010).

%6
\bibitem{Cabra_honeycomb_prb}  D. C. Cabra, C. A. Lamas, and H. D. Rosales, Phys. Rev. B 83, 094506
(2011).

%7
\bibitem{Cabra_honeycomb_2}  D. C. Cabra, C. A. Lamas, and H. D. Rosales,  Modern Physics Letters B (MPLB) 25, 891
(2011).

%8
\bibitem{Zhang_PRB_2013} Hao Zhang and C. A. Lamas, Phys. Rev. B 87, 024415
(2013).


%9
\bibitem{Mulder} A.\ Mulder, R.\ Ganesh, L.\ Capriotti and A.\ Paramekanti, Phys.\
Rev.\ B {\bf 81}, 214419 (2010).

%10
\bibitem{Mosadeq} H. Mosadeq, F. Shahbazi, and S. A. Jafari, J.\ Phys.: Condens.\
Matter {\bf 23}, 226006 (2011).

%11
\bibitem{Albuquerque}
A.~F.~Albuquerque, D.~Schwandt, B.~Het\'{e}nyi, S.~Capponi,
M.~Mambrini, and A.~M.~L\"auchli, Phys.\ Rev.\ B {\bf 84}, 024406
(2011).


%12
\bibitem{Reuther} J. Reuther, D. A. Abanin, and R. Thomale,
\prb{84}{014417}{2011}.

%13
\bibitem{Farnell}
D.~J.~J.~Farnell, R.~F.~Bishop, P.~H.~Y.~Li, J.~Richter, and
C.~E.~Campbell, Phys.\ Rev.\ B {\bf 84}, 012403 (2011).

%14
\bibitem{Bishop_2012} R.~F.~Bishop, P.~H.~Y.~Li, D.~J.~J.~Farnell, and C.~E.~Campbell, J.\ Phys.: Condens.\
Matter {\bf 24}, 236002 (2012).

%15
\bibitem{Li_2012_honeyJ1-J2-J3}
P.~H.~Y.~Li, R.~F.~Bishop, D.~J.~J.~Farnell, and C.~E.~Campbell,
Phys.\ Rev.\ B {\bf 86}, 144404 (2012).

%16
\bibitem{Bishop_2013} R.~F.~Bishop, P.~H.~Y.~Li, and C.~E.~Campbell, J. Phys.: Condens. Matter {\bf 25}, 306002
(2013).

%17
\bibitem{Clark} B.K.\ Clark, D.A.\ Abanin, and  S.L.\ Sondhi Phys. Rev. Lett. {\bf 107}, 087204 (2011).

%18
\bibitem{Mezzacapo} F.~Mezzacapo and M.~Boninsegni, Phys. Rev. B {\bf 85}, 060402(R)
(2012).

%19
\bibitem{Ganesh_PRL_2013} R. Ganesh, J.~van den Brink, and S. Nishimoto,
\prl{110}{127203}{2013}.

%20
\bibitem{Zhu_PRL_2013} Z.~Zhu, D.~A.~Huse, and S.~R.~White,
\prl{110}{127205}{2013}.

%21
\bibitem{Fisher_2013} S. S. Gong, D. N. Sheng, O. I. Motrunich, and M. P. A. Fisher, Phys.\ Rev.\ B {\bf 88}, 165138 (2013).

%22
\bibitem{Albuquerque_capponi_2012} Hong-Yu Yang, A F. Albuquerque, S. Capponi, A. M Lauchli and K. P. Schmidt, New J. Phys. {\bf 14} 115027
(2012).




%%%



%23
\bibitem{BiMnO}
O. Smirnova, M. Azuma, N. Kumada, Y. Kusano, M. Matsuda, Y.
Shimakawa, T. Takei, Y. Yonesaki, and N. Kinomura, J. Am. Chem.
Soc., {\bf 131}, 8313 (2009).


%24
\bibitem{ESR} S.~Okubo, F.~Elmasry, W.~Zhang, M.~Fujisawa, T.~Sakurai, H.~Ohta,
M.~Azuma, O.~A.~Sumirnova, and N.~Kumada,, J.\ Phys.: Conf.\ Series
{\bf 200}, 022042 (2010).

%25
\bibitem{expnew2} M.\ Matsuda, M.\ Azuma, M.\ Tokunaga, Y.\ Shimakawa and N.\ Kumada
Phys.\ Rev.\ Lett.\ {\bf 105}, 187201 (2010).

%26
\bibitem{Azuma} M.~Azuma, M.~Matsuda, N.~Onishi, S.~Olga, Y.~Kusano,
M.~Tokunaga, Y.~Shimakawa, and N.~Kumada, J.\ Phys.: Conf.\ Series
{\bf 320}, 012005 (2011).

%27
\bibitem{Kandpal} H.~C.~Kandpal and J.~van den Brink, Phys. Rev. B {\bf 83}, 140412(R) (2011).


%28
\bibitem{Okumura} S.\ Okumura, H.\ Kawamura, T.\ Okubo, and Y.\ Motome,
J.\ Phys.\ Soc.\ Jpn.\ {\bf 79}, 114705 (2010).

%29
\bibitem{Ganesh_2011} R.\ Ganesh, D.N.\ Sheng, Y.-J.\ Kim and A.\ Paramekanti, Phys.\
Rev.\ B {\bf 83}, 144414 (2011).

%30
\bibitem{Ganesh_QMC} R.\ Ganesh, S.V.\ Isakov, and A.\ Paramekanti, Phys.\
Rev.\ B {\bf 84}, 214412 (2011).

%31
\bibitem{Oitmaa_2012}
J.~Oitmaa and R.~R.~P.~Singh, Phys.\ Rev.\ B {\bf 85}, 014428
(2012).


%%%





%32
\bibitem{Rastelli} E.\ Rastelli, A.\ Tassi, and L.\ Reatto, Physica B {\bf 97}, 1 (1979).

%33
\bibitem{Fouet} J.\ B.\ Fouet, P.\ Sindzingre, and C.\ Lhuillier, Eur.\ Phys.\ J. B {\bf 20}, 241 (2001).

%34
\bibitem{Einarsson} T.\ Einarsson and H.\ Johannesson, Phys.\ Rev.\ B {\bf 43}, 5867 (1991).

%35
\bibitem{Oitmaaj1j2j3} J. Oitmaa and R. R. P. Singh, Phys.\ Rev.\ B {\bf 84}, 094424 (2011).








%36
\bibitem{Trumper1} H.\ A.\ Ceccatto, C.\ J.\ Gazza, and A.\ E.\ Trumper,  Phys.\ Rev.\ B {\bf 47}, 12329 (1993).

%37
\bibitem{Gazza} C.\ J.\ Gazza and H.\ A.\ Ceccato, J.\ Phys.: Condens.\
Matter {\bf 5}, L135 (1993).

%38
\bibitem{Trumper2} A.\ E.\ Trumper, L.\ O.\ Manuel, C.\ J.\ Gazza, and H.\ A.\ Ceccatto,
 Phys.\ Rev.\ Lett.\ {\bf 78}, 2216 (1997).

%39
\bibitem{Coleman} R.\ Flint and P.\ Coleman, Phys.\ Rev.\ B {\bf 79}, 014424 (2009).


%40
\bibitem{Mezio} A.~Mezio, C.~N.~Sposetti, L.~O.~Manuel, and A.~E.~Trumper, Europhys. Lett. {\bf 94}, 47001 (2011).

%41
\bibitem{Feldner} H.~Feldner, D.~C.~Cabra, and G.~L.~Rossini, Phys.\ Rev.\ B {\bf 84}, 214406 (2011).

%42
\bibitem{Messio} L.~Messio, B.~Bernu, and C.~Lhuillier, Phys.\ Rev.\ Lett.\ {\bf 108}, 207204 (2012).

%43
\bibitem{Messio_2013} L.~Messio, C.~Lhuillier, and G.~Misguich, Phys.\ Rev.\ B {\bf 87}, 125127 (2013).

%44
\bibitem{Auerbach} D.\ P.\ Arovas and A.\ Auerbach,  Phys.\ Rev.\ B {\bf 38}, 316
(1988); A.\ Auerbach and D.\ P.\ Arovas, Phys.\ Rev.\ Lett.\ {\bf
61}, 617 (1988).

%45
\bibitem{Auerbach:1994} A.\ Auerbach, {\it Interacting Electrons And Quantum Magnetism}
(Springer-Verlag, New York, 1994).

%46
\bibitem{Auerbach:2011} A.\ Auerbach and D.\ P.\ Arovas, in {\it Introduction to Frustrated Magnetism}, Eds. C.\ Lacroix, P.\ Mendels, and F.\ Mila, (Springer-Verlag, Berlin Heidelberg, 2011).

%47
\bibitem{Mattsson_PRB_1994} A. Mattsson, P. Frojdh, and T. Einarsson, Phys.\ Rev.\ B {\bf 49}, 3997 (1994).

%48
\bibitem{Knetter2000a} C. Knetter and G.S. Uhrig, Eur. Phys. J. B \textbf{13}, 209 (2000).

%49
\bibitem{Wegner1994a} F. Wegner, Ann. Phys. \textbf{506}, 77 (1994).

%50
\bibitem{Arlego2011} M. Arlego and W. Brenig, Phys.\ Rev.\ B {\bf 84}, 134426 (2011).


%%%%%%%%%%------------------------------------
%%%%%%%%%%

%%% refs single layer

% %1
% \bibitem{Bethe} Bethe HA. Z Phys. {\bf 74}, 205 (1931).
%
% %2
% \bibitem{Hamer} Zheng Weihong and C. J. Hamer, Phys.\ Rev.\ B {\bf 47}, 7961 (1993).
%
% %3
% \bibitem{Sanvik_1997} Anders W. Sandvik. Phys.\ Rev.\ B {\bf 56}, 11678 (1997)
%
% %4
% \bibitem{Anderson} P.~W.~Anderson,
% Mater.\ Res.\ Soc.\ Bull.\ {\bf 8}, 153 (1973); P.~Fazekas and
% P.~W.~Anderson, Phil.\ Mag.\ {\bf 30}, 423 (1974); P.~W.~Anderson,
% Science {\bf 235}, 1196 (1987).
% %\bibitem{Anderson} P.\ W.\ Anderson, Science {\bf 235}, 1196 (1987).
%
% %5
% \bibitem{Sachdev1}  M.\ Metlitski and S.\ Sachdev, Phys.\ Rev.\ B {\bf 77},
% 054411 (2008).
%
% %6
% \bibitem{Sachdev2} R. K. Kaul, M. A. Metlitski, S. Sachdev and C. Xu, Phys. Rev. B {\bf 78}, 045110 (2008).
%
% %7
% \bibitem{Sandvik} L.\ Wang and A.\ W.\ Sandvik, Phys.\ Rev.\ B {\bf 81}, 054417
% (2010).
%
% %8
% \bibitem{Moessner} R.\ Moessner , S.L.\ Sondhi and P.\ Chandra, Phys.\ Rev.\ B {\bf 64}, 144416 (2001).
%
% %9
% \bibitem{Poilblanc} A.\ Ralko, M.\ Mambrini and D.\ Poilblanc, Phys.\ Rev.\ B
% {\bf 80}, 184427 (2009).
%
%
% %10
% \bibitem{Mattsson} A.\ Mattsson, P.\ Fr\"{o}jdh and T.\ Einarsson, Phys.\ Rev.\ B {\bf 49}, 3997 (1994).
%
%
% %11
% \bibitem{Takano} K.\ Takano
% Phys.\ Rev.\ B {\bf 74}, 140402 (2006).
%
% %12
% \bibitem{Hermele} M.\ Hermele, Phys.\ Rev.\ B
% {\bf 76}, 035125 (2007).
%
% %13
% \bibitem{Kumar} R.\ Kumar, D.\ Kumar and B.\ Kumar Phys.\
% Rev.\ B {\bf 80}, 214428 (2009).
%
% %14
% \bibitem{BiMnO}
% O. Smirnova, M. Azuma, N. Kumada, Y. Kusano, M. Matsuda, Y.
% Shimakawa, T. Takei, Y. Yonesaki, and N. Kinomura, J. Am. Chem.
% Soc., {\bf 131}, 8313 (2009).
%
% %15
% \bibitem{ESR} S.\ Okubo {\it et al}, J.\ Phys.: Conf.\ Series {\bf 200}, 022042 (2010).
%
% %16
% \bibitem{expnew2} M.\ Matsuda, M.\ Azuma, M.\ Tokunaga, Y.\ Shimakawa and N.\ Kumada
% Phys.\ Rev.\ Lett.\ {\bf 105}, 187201 (2010).
%
% %17
% \bibitem{Azuma} M.\ Azuma {\it et al}, J.\ Phys.: Conf.\ Series {\bf 320}, 012005 (2011).
%
% %18
% \bibitem{libro_experimental} {\it Magnetic Properties of Layered Transition Metal
% Compounds}, Ed.  L.\ J.\ De Jongh, Kluwer, Dordrecht (1990).
%
% %19
% \bibitem{exp_hex2}  A.\ Moller {\it et al}, Phys.\ Rev.\ B {\bf 78}, 024420 (2008).
%
% %20
% \bibitem{expnew1} A.A.\ Tsirlin, O.\ Janson and H.\ Rosner, Phys.\ Rev.\ B {\bf 82}, 144416
% (2010).
%
% %21
% \bibitem{Neto} A.H.\ Castro Neto, F.\ Guinea, N.M.R.\ Peres, K.S.\ Novoselov, and A.K.\ Geim, Rev.\ Mod.\ Phys. {\bf 81}, 109
% (2009).
%
% %22
% \bibitem{HubbardNature} Z.Y.\ Meng, T.C.\ Lang, S.\ Wessel, F.F.\ Assaad, and A.\ Muramatsu,
% Nature {\bf 464}, 847 (2010).
%
% %23
% \bibitem{Wang} F.\ Wang, Phys.\ Rev.\ B {\bf 82}, 024419 (2010).
%
% %24
% \bibitem{Lu} Y.-M.\ Lu and Y.\ Ran, Phys.\ Rev.\ B {\bf 84}, 024420 (2010).
%
% %25
% \bibitem{Yang} H.Y.\ Yang and K.P.\ Schmidt, Europhys. Lett. {\bf 94}, 17004 (2011).
%
% %26
% \bibitem{Vaezi1} A.\ Vaezi and X.G.\ Wen, arXiv:1010.5744v1 [cond-mat.str-el] (2010).
%
% %27
% \bibitem{Vaezi2} A.\ Vaezi, M.\ Mashkoori, and M. Hosseini, Phys.\ Rev.\ B {\bf 85}, 195126 (2012).
%
% %28
% \bibitem{Tran} M.-T.\ Tran and K.-S.\ Kim, Phys.\ Rev.\ B {\bf 83}, 125416 (2011).
%
% %29
% \bibitem{Sorella} S.~Sorella, Y.~Otsuka, and S.~Yunoki arXiv:1207.1783v1 [cond-mat.str-el] (2012).
%
% %30
% \bibitem{Mulder} A.\ Mulder, R.\ Ganesh, L.\ Capriotti and A.\ Paramekanti, Phys.\
% Rev.\ B {\bf 81}, 214419 (2010).
%
% %31
% \bibitem{Ganesh} R.\ Ganesh, D.N.\ Sheng, Y.-J.\ Kim and A.\ Paramekanti, Phys.\
% Rev.\ B {\bf 83}, 144414 (2011).
%
% %32
% \bibitem{Okumura} S.\ Okumura, H.\ Kawamura, T.\ Okubo and Y.\ Motome,
% J.\ Phys.\ Soc.\ Jpn.\ {\bf 79}, 114705 (2010).
%
% %33
% \bibitem{Clark} B.K.\ Clark, D.A.\ Abanin and  S.L.\ Sondhi Phys. Rev. Lett. {\bf 107}, 087204 (2011).
%
% %34
% \bibitem{Mosadeq} H. Mosadeq, F. Shahbazi, and S. A. Jafari, J.\ Phys.: Condens.\
% Matter {\bf 23}, 226006 (2011).
%
% %35
% \bibitem{Cabra2012} H. D. Rosales, D. C. Cabra, C. A. Lamas, P. Pujol, M. E. Zhitomirsky, arXiv:1208.2416 (2012)
%
% %36
% \bibitem{Mezzacapo} F.~Mezzacapo and M.~Boninsegni, Phys. Rev. B {\bf 85}, 060402(R)
% (2012).
%
% %37
% \bibitem{Bishop} R.~F.~Bishop, P.~H.~Y.~Li, D.~J.~J.~Farnell, and C.~E.~Campbell J.\ Phys.: Condens.\
% Matter {\bf 24}, 236002 (2012).
%
% %38
% \bibitem{nos} D.C.\ Cabra, C.A.\ Lamas, and H.D.\ Rosales, Phys.\ Rev.\ B {\bf 83}, 094506 (2011).
%
% %39
% \bibitem{Albuquerque}
% A.~F.~Albuquerque, D.~Schwandt, B.~Het\'{e}nyi, S.~Capponi,
% M.~Mambrini, and A.~M.~L\"auchli, Phys.\ Rev.\ B {\bf 84}, 024406
% (2011).
%
% %40
% \bibitem{Oitmaa}
% J.~Oitmaa and R.~R.~P.~Singh, Phys.\ Rev.\ B {\bf 84}, 094424
% (2011).
%
% %41
% \bibitem{Farnell}
% D.~J.~J.~Farnell, R.~F.~Bishop, P.~H.~Y.~Li, J.~Richter, and
% C.~E.~Campbell, Phys.\ Rev.\ B {\bf 84}, 012403 (2011).
%
% %42
% \bibitem{Reuther}
% J.~Reuther, D.~A.~Abanin, and R.~Thomale, Phys.\ Rev.\ B {\bf 84},
% 014417 (2011).
%
% %43
% \bibitem{Li_honeycomb_J1neg}
% P.~H.~Y.~Li, R.~F.~Bishop, D.~J.~J.~Farnell, J.~Richter, and
% C.~E.~Campbell, Phys.\ Rev.\ B {\bf 85}, 085115 (2012).
%
% %44
% \bibitem{Bishop_PRB} R.~F.~Bishop and P.~H.~Y.~Li, Phys.\ Rev.\ B {\bf 85}, 155135 (2012).
%
% %45
% \bibitem{Li_2012_honeyJ1-J2-J3}
% P.~H.~Y.~Li, R.~F.~Bishop, D.~J.~J.~Farnell, and C.~E.~Campbell,
% arXiv:1204.2390v1 [cond-mat.str-el] (2012).
%
% %46
% \bibitem{Wen_1991} X.\ G.\ Wen , Phys.\ Rev.\ B {\bf 44}, 2664 (1991).
%
% %47
% \bibitem{LM} C.\ Lhuillier and G.\ Misguich, in {\it Introduction to Frustrated Magnetism}, Eds. C.\ Lacroix, P.\ Mendels, and F.\ Mila, (Springer-Verlag, Berlin Heidelberg, 2011).
%
% %48
% \bibitem{Katsura} S.\ Katsura, T.\ Ide, and T.\ Morita, J.\ Stat.\
% Phys. {\bf 42}, 381 (1986)
%
% %49
% \bibitem{Rastelli} E.\ Rastelli, A.\ Tassi, and L.\ Reatto, Physica B {\bf 97}, 1 (1979).
%
% %50
% \bibitem{Fouet} J.\ B.\ Fouet, P.\ Sindzingre and C.\ Lhuillier, Eur.\ Phys.\ J. B {\bf 20}, 241 (2001).
%
% %51
% \bibitem{Einarsson} T.\ Einarsson and H.\ Johannesson, Phys.\ Rev.\ B {\bf 43}, 5867 (1991).
%
% %52
% \bibitem{Coleman} R.\ Flint and P.\ Coleman, Phys.\ Rev.\ B {\bf 79}, 014424 (2009).
%
% %53
% \bibitem{Trumper1} H.\ A.\ Ceccato, C.\ J.\ Gazza and A.\ E.\ Trumper,  Phys.\ Rev.\ B {\bf 47}, 12329 (1993).
%
% %54
% \bibitem{Gazza} C.\ J.\ Gazza and H.\ A.\ Ceccato, J.\ Phys.: Condens.\
% Matter {\bf 5}, L135 (1993).
%
% %55
% \bibitem{Trumper2} A.\ E.\ Trumper, L.\ O.\ Manuel, C.\ J.\ Gazza and H.\ A.\ Ceccatto,
%  Phys.\ Rev.\ Lett.\ {\bf 78}, 2216 (1997).
%
% %56
% \bibitem{Mezio} A.~Mezio, C.~N.~Sposetti, L.~O.~Manuel and A.~E.~Trumper, Europhys. Lett. {\bf 94}, 47001 (2011).
%
% %57
% \bibitem{Feldner} H.~Feldner, D.~C.~Cabra, and G.~L.~Rossini, Phys.\ Rev.\ B {\bf 84}, 214406 (2011).
%
% %58
% \bibitem{Messio} L.~Messio, B.~Bernu, and C.~Lhuillier, Phys.\ Rev.\ Lett.\ {\bf 108}, 207204 (2012).
%
% %59
% \bibitem{Auerbach} D.\ P.\ Arovas and A.\ Auerbach,  Phys.\ Rev.\ B {\bf 38}, 316
% (1988); A.\ Auerbach and D.\ P.\ Arovas, Phys.\ Rev.\ Lett.\ {\bf
% 61}, 617 (1988).
%
% %60
% \bibitem{Auerbach:1994} A.\ Auerbach, {\it Interacting Electrons And Quantum Magnetism}
% (Springer-Verlag, New York, 1994).
%
% %61
% \bibitem{Auerbach:2011} A.\ Auerbach and D.\ P.\ Arovas, in {\it Introduction to Frustrated Magnetism}, Eds. C.\ Lacroix, P.\ Mendels, and F.\ Mila, (Springer-Verlag, Berlin Heidelberg, 2011).
%
% %62
% \bibitem{Hirsch} J.\ E.\ Hirsch and Sanyee.\ Tang,  Phys.\ Rev.\ B {\bf 39},
% 2850 (1989)
%
%
% %---------------------------------------------------------
% %---------------------------------------------------------
%
%
%
% %63
% \bibitem{Colpa} J.~H.~P.~Colpa, Physica A {\bf 93}, 327 (1978).
%
% %64
% \bibitem{Zheng} Zheng Weihong, J. Oitmaa, and C. J. Hamer, Phys.\ Rev.\ B {\bf 44}, 11869 (1991).
%
% %65
% \bibitem{White_1992} S.~R.~White, Phys.\ Rev.\ Lett.\ {\bf 69}, 2863 (1992).
%
% %66
% \bibitem{White_2007} S.~R.~White and A.~L.~Chernyshev, Phys.\ Rev.\ Lett.\ {\bf 99}, 127004 (2007).
%
% %67
% \bibitem{Stoudenmire} E.~M.~Stoudenmire and S.~R.~White, Annual Review of Condensed Matter Physics {\bf 3}, 111 (2012).
%
% %68
% \bibitem{Yan} S.~M.~Yan, D.~A.~Huse and S.~R.~White, Science {\bf 332}, 1173 (2011).
%
% %69
% \bibitem{Depenbrock} S.~Depenbrock, I.~P.~McCulloch and U.~Schollw{\"o}ck, arXiv:1205.4858v1 [cond-mat.str-el] (2012).
%
% %70
% \bibitem{Jiang} H.~C.~Jiang, H.~Yao, and L.~Balents, Phys.\ Rev.\ B {\bf 86}, 024424 (2012).
%
%
%
%
%
%
%
%
%
%

% \bibitem{Matsuda_prl_2010} M. Matsuda, M. Azuma, M. Tokunaga, Y. Shimakawa, and N. Kumada, \prl{105}{187201}{2010}
%
% \bibitem{Matsuda_2011} M. Azuma, M. Matsuda, N. Onishi, S. Olga, Y. Kusano,  M. Tokunaga, Y. Shimakawa, and N. Kumada,
%  Journal of Physics: Conference Series {\bf 320}, 012005 (2011)
%
%
% \bibitem{spin_1_dimerization}  Yu-Wen\  Lee, Min-Fong\  Yang,  arXiv:1110.0962
%
% \bibitem{spin_1_plaquette}  H. H. Zhao, Cenke Xu, Q. N. Chen, Z. C. Wei, M. P. Qin, G. M. Zhang, T. Xiang,  arXiv:1105.2716
%
% \bibitem{Meng_Muramatsu_Nature} Z. Y.\ Meng, T. C.\ Lang, S.\ Wessel, F. F.\ Assaad and A.Muramatsu, \nature{464}{847}{2010}
%
%
% \bibitem{Mila_triangular_biquadratic} A.\ Laeuchli, F.\ Mila, K.\ Penc, \prl{97}{087205}{2006}
%
% \bibitem{Rastelli} E.\ Rastelli, A.\ Tassi, L.\ Reatto, Physica 97B, 1 (1979).
%
% \bibitem{Fouet} J.B.\ Fouet, P.\ Sindzingre, C.\ Lhuillier, Eur.\ Phys.\ J. B {\bf 20}, 241 (2001)




%%%%%%%%%%%%%%%%%%%%%%%%%%%%%%%%%%%%%%%%%%%%%%%%%%%%%%%%%%%%%%%%%%%%%%%%%%%%%%%%%%%%%%%%%%%%%%%%%%%%%%%%%%%%%%
%%%%%%%%%%%%%%%%%%%%%%%%%%%%%%%%%%%%%%%%%%%%%%%%%%%%%%%%%%%%%%%%%%%%%%%%%%%%%%%%%%%%%%%%%%%%%%%%%%%%%%%%%%%%%%
%%%%%%%%%%%%%%%%%%%%%%%%%%%%%%%%%%%%%%%%%%%%%%%%%%%%%%%%%%%%%%%%%%%%%%%%%%%%%%%%%%%%%%%%%%%%%%%%%%%%%%%%%%%%%%
%%%%%%%%%%%%%%%%%%%%%%%%%%%%%%%%%%%%%%%%%%%%%%%%%%%%%%%%%%%%%%%%%%%%%%%%%%%%%%%%%%%%%%%%%%%%%%%%%%%%%%%%%%%%%%
%%%%%%%%%%%%%%%%%%%%%%%%%%%%%%%%%%%%%%%%%%%%%%%%%%%%%%%%%%%%%%%%%%%%%%%%%%%%%%%%%%%%%%%%%%%%%%%%%%%%%%%%%%%%%%
%%%%%%%%%%%%%%%%%%%%%%%%%%%%%%%%%%%%%%%%%%%%%%%%%%%%%%%%%%%%%%%%%%%%%%%%%%%%%%%%%%%%%%%%%%%%%%%%%%%%%%%%%%%%%%
%%%%%%%%%%%%%%%%%%%%%%%%%%%%%%%%%%%%%%%%%%%%%%%%%%%%%%%%%%%%%%%%%%%%%%%%%%%%%%%%%%%%%%%%%%%%%%%%%%%%%%%%%%%%%%
%%%%%%%%%%%%%%%%%%%%%%%%%%%%%%%%%%%%%%%%%%%%%%%%%%%%%%%%%%%%%%%%%%%%%%%%%%%%%%%%%%%%%%%%%%%%%%%%%%%%%%%%%%%%%%



% \bibitem{Cabra1} D. C.\ Cabra, C. A.\ Lamas, H. D.\ Rosales, \prb{83}{094506}{2011}
%
% \bibitem{Cabra2} D. C.\ Cabra, C. A.\ Lamas, H. D.\ Rosales, International Journal of
%  Modern Physics B {\bf 25}, 891 (2011)



% \bibitem{Cabra_honey_prb_2011} XXX
%
% \bibitem{Cabra_honey_mplb_2011} xxx
%
% \bibitem{Sachdev1}  M.\ Metlitski, S.\ Sachdev, Phys.\ Rev.\ B {\bf 77},
% 054411 (2008).
%
% \bibitem{Sachdev2} R.\ Kaul {\it et al.},
% Phys.\ Rev.\ B {\bf 78}, 045110 (2008).
%
% \bibitem{Sandvik} L.\ Wang, A.W.\ Sandvik, Phys.\ Rev.\ B {\bf 81}, 054417
% (2010).
%
% \bibitem{Poilblanc} A.\ Ralko, M.\ Mambrini, D.\ Poilblanc, Phys.\ Rev.\ B
% {\bf 80}, 184427 (2009).
%
% \bibitem{Anderson} P.W.\ Anderson, Science {\bf 235}, 1196 (1987).
%
% \bibitem{ESR} S.\ Okubo {\it et al}, J.\ Phys.: Conf.\ Series {\bf 200}, 022042 (2010).
%
% \bibitem{libro_experimental} {\it Magnetic Properties of Layered Transition Metal
% Compounds}, Ed.  L.J.\ De Jongh, Kluwer, Dordrecht (1990).
%
% \bibitem{design}
% R.\ Arita {\it et al.}, Phys.\ Rev.\ Lett.\ {\bf 88}, 127202 (2002);
% Y.\ Suwa {\it et al.}, Phys.\ Rev.\ B {\bf 68}, 174419 (2003);
% A.\ Harrison, J. Phys.: Condens.\ Matter {\bf 16}, S553 (2004);
% Y.\ Z.\ Zheng {\it et al.}, Angew.\ Chem.\ Int.\ Ed.\ {\bf 46}, 6076 (2007).
%
% \bibitem{cold_atom_a}
% D.\ Jaksch, P.\ Zoller, Ann.\ Phys.\ (N.Y.) {\bf 315}, 52 (2005).
%
% \bibitem{cold_atoms}
% M.\ Lewenstein {\it et al.}, Adv.\ Phys.\ {\bf 56}, 243 (2007).
%

%

%
% \bibitem{square} P.\ Chandra, B.\ Doucot, Phys.\ Rev.\ B 38, 9335 (1988);
% E. Dagotto and A. Moreo, Phys. Rev. Lett. 63, 2148
% (1989); N. Read and S. Sachdev, Phys. Rev. Lett. 66,
% 1773 (1991); H.J. Schulz and T.A.L. Ziman, Europhys.
% Lett. 18, 355 (1992); F. Figueirido et al ., Phys. Rev. B
% 41, 4619 (1990); M.E. Zhitomirsky and K. Ueda, ibid. 54
% 9007 (1996); R.R.P. Singh et al ., ibid. 60, 7278 (1999);
% V.N. Kotov et al ., ibid. 60, 14613 (1999); L. Capriotti
% and S. Sorella, Phys. Rev. Lett. 84, 3173 (2000).
%
%

%\bibitem{Takahashi} M. Takahashi, J.\ Phys.\ C  {\bf 10}, 1289 (1977).

%\bibitem{MacDonald} A.H.\  MacDonald {\it et al.}. Phys.\ Rev.\ B {\bf 37}, 9753 (1988).

%\bibitem{Sorella} S. Sorella, E.\ Tosatti, Eur.\ Phys.\ Lett.\  {\bf 19}, 699-704 (1992).

%\bibitem{Lamas} C.A.\ Lamas, D.C.\ Cabra, N.\ Grandi, Phys.\ Rev.\ B {\bf 80}, 075108 (2009).

%\bibitem{Feldner} H.\ Feldner {\it et al.},  Phys.\ Rev.\ B {\bf 81}, 115416 (2010).

%\bibitem{Sun} G.Y.\ Sun, S.P.\ Kou, ArXiv:0911.3002 (2009).

%\bibitem{LesHouches} Z.Y.\ Meng, Private Comunication.


% \bibitem{Auerbach} A.\ Auerbach, {\it Interacting Electrons And Quantum Magnetism}
% (Springer-Verlag, New York, 1994).
%
% \bibitem{aclaracion} The $J_1-J_2-J_3$ Heisenberg model on the honeycomb lattice was analyzed previously using
% LSWT in \cite{Fouet}, but the intermediate phase along the line $J_2 = J_3$, which is the main aim of the present
% paper, was not analyzed in detail.
%
% \bibitem{Spinwave1} A.\ Altland and B.\ Simons,  {\it Condensed Matter Field Theory}
% (Cambridge University Press, Cambridge, England, 2006).
%
% \bibitem{Spinwave2} A.E.\ Trumper, L.\ Capriotti, S.\ Sorella, Phys.\ Rev.\ B {\bf 61}, 11529 (1999).
%
% \bibitem{Matson} A.\ Mattson, P.\ Frojdh, Phys.\ Rev.\ B {\bf 49}, 3997 (1994).
%
% \bibitem{Trumper1} H.A.\ Ceccato, C.J.\ Gazza, A.E.\ Trumper,  Phys.\ Rev.\ B {\bf 47}, 12329 (1993).
%
% \bibitem{Coleman} R.\ Flint {\it et al.}, Phys.\ Rev.\ B {\bf 79}, (2008).
%
% \bibitem{Colpa} J.H.P.\ Colpa, Physica 93A, 327-353 (1978)
%
% \bibitem{Trumper2} A.E.\ Trumper {\it et al.}, Phys.\ Rev.\ Lett.\ {\bf 78}, 2216 (1997)
%
% \bibitem{spinpack}  J.\ Schulenburg, program package SPINPACK, http://www-e.uni-
% magdeburg.de/jschulen/spin/

\end{thebibliography}
\end{document}